\documentclass[a4paper]{jpconf}

\usepackage[utf8]{inputenc} % allow utf-8 input
\usepackage[T1]{fontenc}    % use 8-bit T1 fonts
\usepackage[colorlinks=true,allcolors=blue]{hyperref}       % hyperlinks
\usepackage{url}            % simple URL typesetting
\usepackage{booktabs}       % professional-quality tables
\usepackage{amsfonts}       % blackboard math symbols
\usepackage{nicefrac}       % compact symbols for 1/2, etc.
\usepackage{microtype}      % microtypography
\usepackage{graphicx}
\usepackage{wrapfig}
\usepackage[toc,page]{appendix}
\usepackage{subcaption}
\usepackage{todonotes}
\usepackage{xcolor}
\usepackage{fancyvrb}
\usepackage{textcomp}

\newcommand{\madjax}{\texttt{MadJax}}
\newcommand{\madgraph}{\texttt{MadGraph}}
\newcommand{\jax}{\texttt{JAX}}

\definecolor{myGreen}{HTML}{026C16}
\definecolor{myBlue}{HTML}{0025FF}
\definecolor{myRed}{HTML}{BA2421}
\definecolor{myGray}{HTML}{676767}
\definecolor{myTeal}{HTML}{40807F}

\begin{document}

\title{Differentiable Matrix Elements with \madjax{}}

\author{Lukas Heinrich}
\address{Technical University of Munich}
\ead{lukas.heinrich@cern.ch}
\author{Michael Kagan}
\address{SLAC National Accelerator Laboratory}
\ead{makagan@slac.stanford.edu}

\begin{abstract}
   \madjax~is a tool for generating and evaluating differentiable matrix elements of high energy scattering processes. As such, it is a step towards a differentiable programming paradigm in high energy physics that facilitates the incorporation of high energy physics domain knowledge, encoded in simulation software, into gradient based learning and optimization pipelines. \madjax~comprises two components: (a) a plugin to the general purpose matrix element generator \madgraph~that integrates matrix element and phase space sampling code with the \jax~differentiable programming framework, and (b) a standalone wrapping API for accessing the matrix element code and its gradients, which are computed with automatic differentiation. The \madjax~implementation and example applications of simulation based inference and normalizing flow based matrix element modeling, with capabilities enabled uniquely with differentiable matrix elements, are presented.
\end{abstract}

\section{Introduction}

Matrix Element (ME) Monte Carlo generators provide a theoretical description of differential cross sections for high-energy particle scattering processes encapsulated in software. Theoretical calculations are encoded in stochactic simulators to allow for Monte Carlo integration and sampling of scattered particle configurations.   Together with parton-shower simulations of QCD radiation processes, and models of hadronization, these programs can be used to sample particle interactions and final state particle kinematic configurations for the experimental analysis of collider data, such as data from the LHC for future colliders. However, while collider physics data analysis typically relies on large simulated datasets for statistical inference, simulators do not retain event-by-event information on the dependence of simulated predictions on input parameters or variables. Ultimately, this limits the ways in which researchers can use these simulators.

With \madjax, access to the dependence of simulated predictions on inputs is enabled through gradients. \madjax~is a differentiable programming (DP)~\cite{NEURIPS2018_34e15776} approach to the simulation of matrix elements, which combines the python-based code generation of the \madgraph~matrix element generator~\cite{madgraph2011} with the \jax~DP framework~\cite{jax2018github}. In DP, software is written in, or transformed into, differentiable code via the use of automatic differentiation (AD)~\cite{autodiff}, an algorithmic way to efficiently evaluate derivatives of computer programs. This approach is flexible and optimizable; differentiable HEP software and machine learning (ML) tools can be mixed, for instance to use ML surrogates of non-differentiable computations, and can be jointly optimized  to improve speed and prediction accuracy. HEP simulation tools, and the knowledge they encode, can be used as physics prediction engines directly within ML pipelines for developing physics-informed ML tools.

\begin{wrapfigure}{r}{0.45\textwidth}
    \centering
    \vspace{-0.5cm}
    \includegraphics[width=0.45\textwidth]{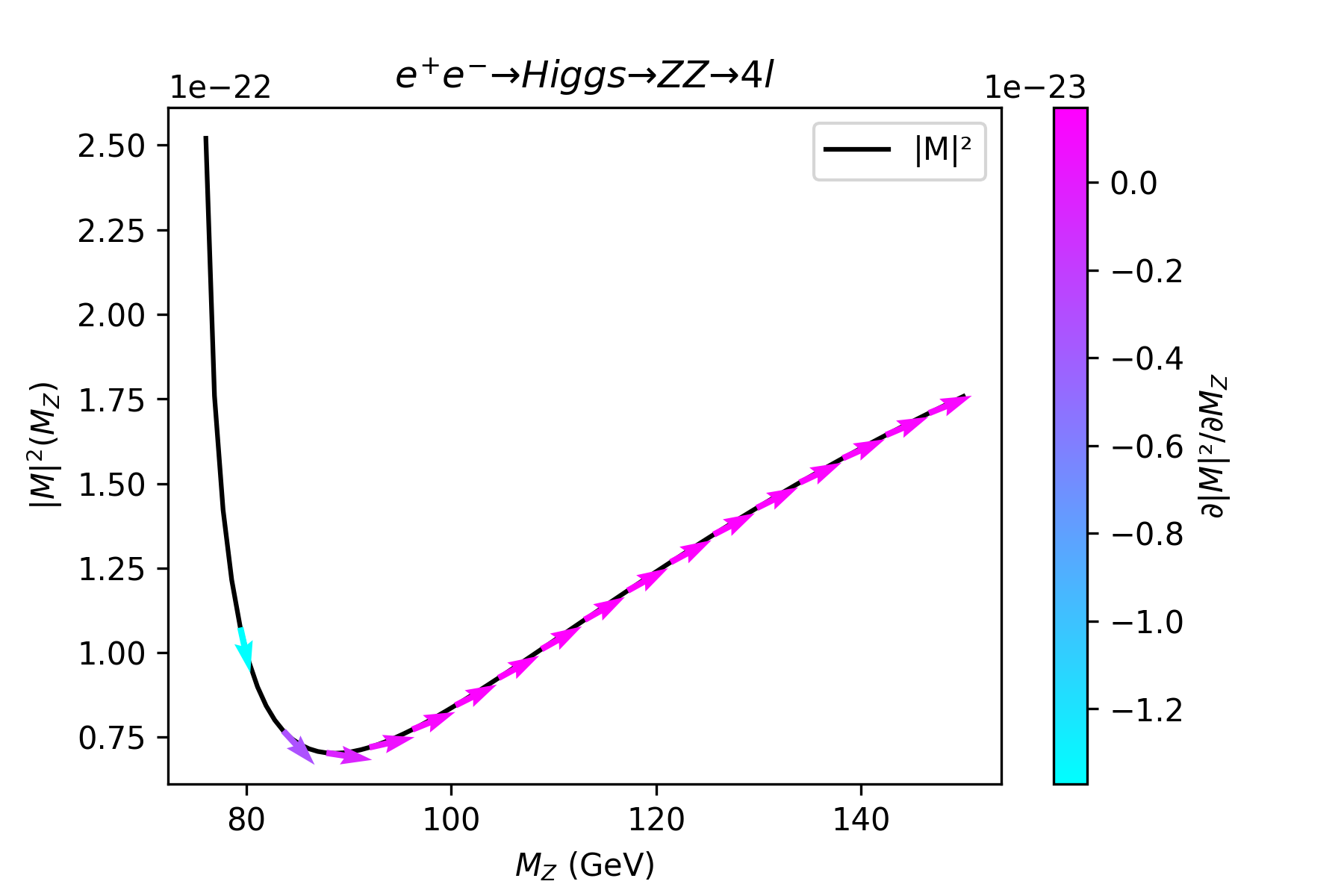}
    \vspace{-0.4cm}
    \caption{Matrix element $\mathcal{M}(r|M_Z)$ for the process $e^{+}e^{-}\to H \to ZZ^(*)\to ee\mu\mu$ and its gradients evaluated by \madjax{} at the phase space point $r = [0.2]^8$}
    \label{fig:h4l}
    \vspace{-0.3cm}
\end{wrapfigure}

\madjax~enables the evaluation of MEs and their gradients automatically. \madjax~also includes a differentiable implementation of the RAMBO algorithm~\cite{platzer2013rambo}, which facilitates transforming particle configurations into \textit{phase space} points on a hypercube for aiding sampling algorithms. An example of the $e^{+}e^{-}\to h \to ZZ\to 4l$ matrix element and \madjax-computed gradients with respect to the $Z$ boson mass is seen in Fig.~\ref{fig:h4l}. To the best of our knowledge, \madjax~is the first differentiable matrix element generator.

Related work on AD / DP in physics are discussed below. Sec.~\ref{sec:imp} presents the implementation, usage, and performance of \madjax{}. Sec.~\ref{sec:app} presents applications of \madjax{} in case studies of simulation-based inference and ME modeling with ML.

\subsection{Related Work}

Automatic differentiation~\cite{autodiff,JMLR:v18:17-468}, and its use in gradient based optimization, is ubiquitous in ML, statistics, and applied math. AD uses the chain rule to evaluate derivatives of a function that is represented as a computer program.  AD takes as input program code and produces new code for evaluating the program and derivatives. AD typically builds a computational graph, or a directed acyclic graph of mathematical operations applied to an input. Gradients are defined for each operation of the graph, and the total gradient can be evaluated from input to output, called forward mode, and from output to input, called reverse mode or backpropagation in ML. AD is backbone of ML / DP frameworks like \texttt{TensorFlow}~\cite{tensorflow2015-whitepaper}, \jax~\cite{jax2018github}, and \texttt{PyTorch}~\cite{NEURIPS2019_9015pytorch}.

In HEP, AD has been used in histogram fitting in \texttt{pyHF}~\cite{pyhf, pyhf_joss}, and \texttt{Neos}~\cite{lukas_heinrich_2020_3697981} builds on \texttt{pyHF} to enable joint histogram fitting and analysis optimization. For modeling parton distribution functions (pdf) used by matrix element generators, Ref.~\cite{pdflow2021} implemented the NNPDF software in TensorFlow to enable to fast evaluation on GPUs. More recently, fitting of neural network pdf models implemented in \texttt{TensorFlow} has been explored by the NNPDF collaboration~\cite{ball2021opensource}. Concurrent with our work, \texttt{MadFlow}~\cite{carrazza2021madflow} has explored using a \texttt{TensorFlow} backend for \madgraph{} for GPU acceleration enabled but did not explore AD capabilities. 
 
 Outside of HEP, the development of differentiable physics simulators is an active area of research in ML and scientific computing. The growing body of examples includes: Rigid body dynamics ~\cite{todorov2012mujoco,Lee2018,10.1145/2776880.2792704,NEURIPS2018_842424a1,QiaoLKL20,hu2019difftaichi} and soft-body~\cite{hu2019chainqueen,huang2021plasticinelab,NEURIPS2019_28f0b864} simulators have been used in reinforcement learning, robotics, optimization, and control problems. Differentiable light transport and Monte Carlo ray tracing is used in rendering and graphics~\cite{Li:2018:DMC,bangaru2020warpedsampling,Zeltner2021MonteCarlo}
 including neural rendering work like NeRF~\cite{mildenhall2020nerf}. Differentiable simulation of fluid dynamics, for prediction, control, and optimizing solutions of PDEs like Navier-Stokes, are an active research area~\cite{Holl2020Learning,Ummenhofer2020Lagrangian,NEURIPS2020_43e4e6a6}, as are differentiable molecular dynamics simulations~\cite{jaxmd2020,doerr2020torchmd} for computational condensed matter physics.

\section{Implementation}\label{sec:imp}

\madgraph{} is a tool for the computation of matrix elements and event generation at the parton level and is widely used at the LHC. As a general purpose tool, \madgraph{} receives as input a given physics transition, i.e. the desired input and output particles, and dynamically compiles a list of hard processes for which matrix elements $\mathcal{M}(x)$ need to be computed. It then generates code for matrix elements in several possible target languages. The most commonly used ones are \texttt{C++} and \texttt{FORTRAN} due to their high evaluation performance. A \texttt{Python} output option is also available, but its performance as a dynamically interpreted language is much reduced. The Matrix Elements are themselves calculated based on helicity amplitudes as generated by ALHOA~\cite{DEAQUINO20122254}.

In \madjax, integration with \jax~is performed during \madgraph~automated matrix element code generation. To do so, several challenges had to be addressed, including:
(a) control flow, such as conditional and loop statements, required
adaptation to allowable control flow paradigms for the \texttt{LAX} backend of \jax{}, (b) some components required adaptation to the functional programming paradigm of \jax{}, and (c) some numerical methods required custom derivative implementation, such as implicit differentiation for root-finding. Code vectorization with \jax's~\texttt{vmap\{...\}} functionality was use to increase computational speed and efficiency. For example, loop statements, such as loops over helicity states, were vectorized where possible. Multi-event phase space sampling, and ME and ME gradient evaluation, were enabled with vectorization. \madjax{} inherits both CPU and GPU compatibility from \jax. However, many of the computations in \madjax{} generated code do not have optimized GPU kernels and GPU evaluation did not result in significant speed-up.

\subsection{Usage}

\madjax{} offers an easy-to-use API and transparently connects to the plugin mechanism of \madgraph{}. It comprises two separate sub-packages: (1) a standard \madgraph{} plugin to generate matrix element code in the \jax{} language and must be run within the \madgraph{} environment, and (2) a standalone wrapping API around the generated code which can be used independently from \madgraph{}. The first stage is achieved by adding the \madjax~plugin to the \verb+PLUGINS+ directory of \madgraph{} and requesting the \madjax{} output in the run card via  \verb+mg5_aMC --mode=madjax_me_gen run.mg5+. This step produces a directory with generated, differentiable \madjax{} code, which is then accessible through the \madjax{} python API. An example listing is shown in Figure~\ref{fig:mj_snippet}.
\begin{figure}
1. Generation:

\begin{Verbatim}[commandchars=\\\{\}]
generate p p > t t~, t > b udsc udscx , t~ > b~ udsc udscx
output madjax generated_ttbar
\textcolor{myGreen}{set} auto_update 0
\end{Verbatim}

2. Evaluation:
\begin{Verbatim}[commandchars=\\\{\}]
\textcolor{myGreen}{import} \textcolor{myBlue}{madjax}
mj = madjax\textcolor{myGray}{.}MadJax(\textcolor{myRed}{\textquotesingle{}generated_ttbar\textquotesingle})
E_cm = \textcolor{myGray}{14000} \textcolor{myTeal}{\textit{#GeV}}
process = \textcolor{myRed}{\textquotesingle{}Matrix_1_gg_ttx_t_budx_tx_bxdux\textquotesingle}
matrix_element = mj\textcolor{myGray}{.}matrix_element(E_cm,process)

parameters = {(\textcolor{myRed}{\textquotesingle{}mass\textquotesingle},\textcolor{myGray}{6}): \textcolor{myGray}{173.0}} \textcolor{myTeal}{\textit{#set top mass}}
phasespace_coords = [\textcolor{myGray}{0.1}]\textcolor{myGray}{*14} \textcolor{myTeal}{\textit{#14D phasespace}}

val, grad = matrix_element(parameters,phasespace_coords)
grad[(\textcolor{myRed}{\textquotesingle{}mass\textquotesingle}, \textcolor{myGray}{6})] \textcolor{myTeal}{\textit{#gradient wrt top mass}}
\end{Verbatim}
    \caption{Example Usage of the \madjax{} code generation and evaluation API}
    \label{fig:mj_snippet}
\end{figure}

The ME function allows the user to explicitly pass parameter values through a mapping from SLHA block identifiers to  parameter values (here: top quark mass). By default the \madjax{} returns gradients with respect to the parameters, but gradients with respect to the phase-space point can be requested as well. The phasespace point is identified through a set of hypercube coordinates.

\begin{table}[t]
    \centering
    \begin{tabular}{c|c|c}
         \toprule
         Evaluation Mode & Performance & Output \\\midrule
         \jax{} no-JIT &  5.1s / evaluation & value and gradient \\
         \jax{} JIT &  15.9s / 100k evaluations & value and gradient\\
         \jax{} JIT + \texttt{vmap} & 5.6s / 100k evaluations & value and gradient\\
         Standalone C++ & 0.8s / 100k evaluations & value only \\
         \bottomrule
    \end{tabular}
    \caption{Performance Evaluations for a top quark decay example process}
    \label{tab:performance}
    \vspace{-0.6cm}
\end{table}

\subsection{Performance}

While the implementation of \madjax{} makes heavy use of the \texttt{Python} export option, its performance is significantly better due to the use of built in just-in-time (JIT) compilation features of \jax{}. In Table \ref{tab:performance} we report performance results for the gluon-initiated ttbar production matrix element (\verb+Matrix_1_gg_ttx_t_budx_tx_bxdux+), in which a computation of the ME value and its gradients via reverse-mode differentiation stays within an order or magnitude of the standalone C++ code the MadGraph package produces. Here, the JIT compilation and vectorization (\texttt{vmap}) capabilities of \jax{} provide a over 10,000x speedup over the non-compiled \jax{} code.

\begin{wrapfigure}{r}{0.45\textwidth}
    \centering
    \vspace{-0.85cm}
    \includegraphics[width=0.45\textwidth]{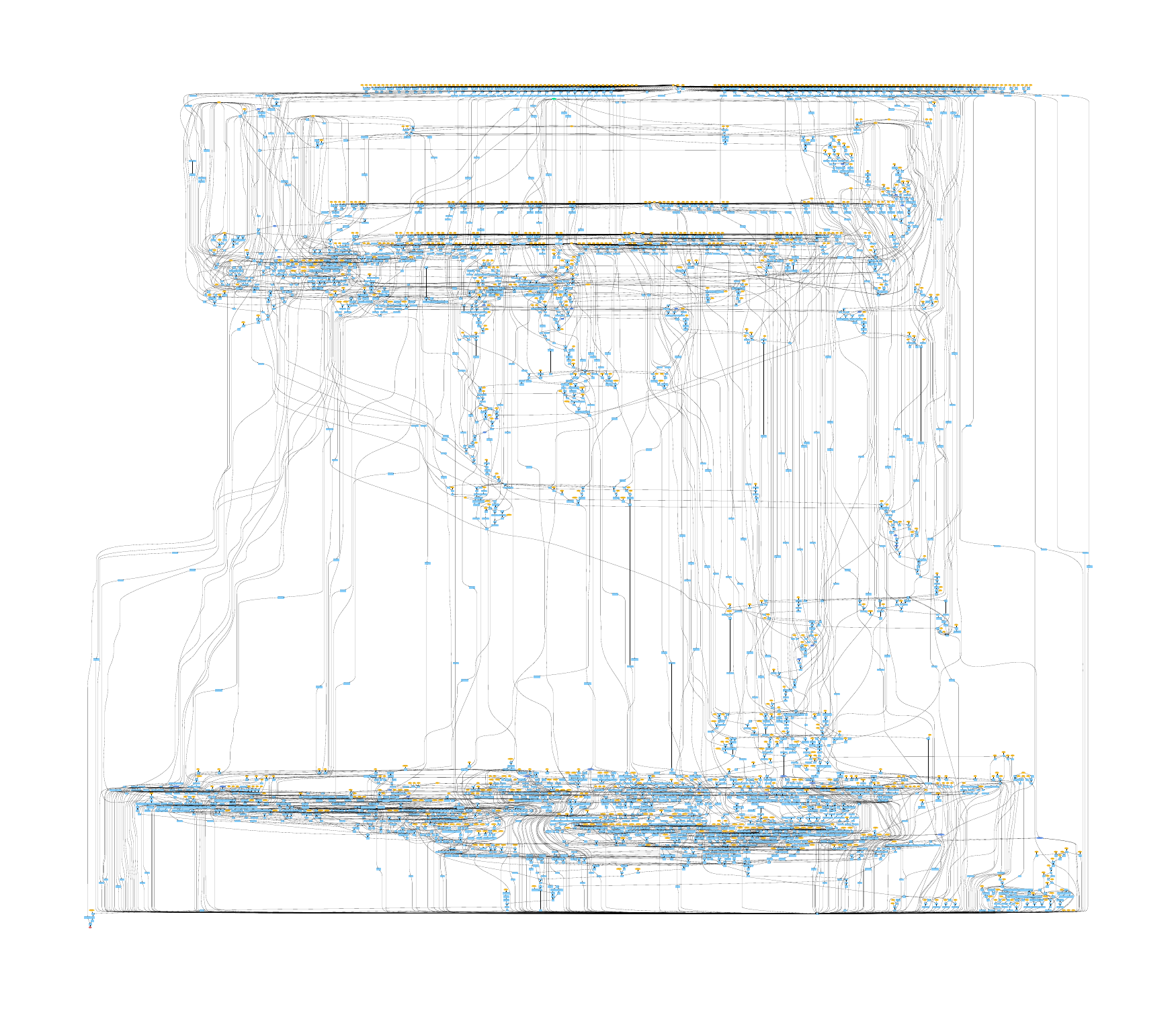}
    \vspace{-0.9cm}
    \caption{Compute graph generated by \madjax{} for the $e^{+}e^{-} \to t\bar{t} \to 6q$ process }
    \label{fig:mj_viz}
    \vspace{-1cm}
\end{wrapfigure}
Several challenges were observed when using \jax{}~for AD integration. Most notably, the \madgraph{}~generated code can contain control flow such as branching conditionals and loop statements which are known to cause inefficiencies during compute graph generation in \jax{}~\cite{jaxSharpBits}. This can be seen in Fig.~\ref{fig:mj_viz} where \jax{}~generates a massive and complex compute graph even for the few hundred lines of code describing the $e^{+}e^{-}\to t\bar{t} \to q\bar{q} b q\bar{q} \bar{b}$ process. The python code generated by \madgraph{} is a near direct translation from the original \texttt{FORTRAN} code, and is not well adapted for JIT compile time code optimization in \jax{}. As a result, even the few hundred lines of code generated by \madjax{} for this process could take 10min or longer to compile.

\section{Applications}\label{sec:app}

Two case studies are presented below that show how the use of ME gradients can improve HEP data analysis techniques. As MEs evaluate the differential cross section of a given particle configuration, they can be interpret as un-normalized\footnote{The total cross section would be the normalizing constant} probability density functions $\tilde{p}(x)$. With \madjax, the gradients $\nabla_{x} \tilde{p}(x)$ can be computed, as can gradients of RAMBO transformations.

\subsection{Mining Gold with Differentiable Simulators}

\begin{figure}[ht]
  \begin{center}
\includegraphics[width=\textwidth]{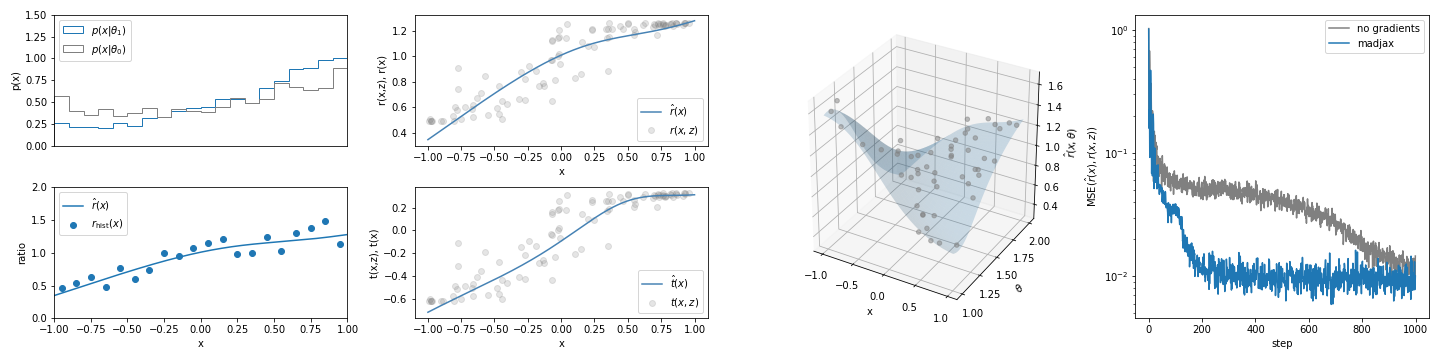}
%\vspace{-0.2cm}
\caption{SBI example using \madjax~for learning
  ML based likelihood ratios as a function of the theory parameter
  $G_F$ in $e^{+} e^{-} \to Z \to \mu^{+} \mu^{-} $ . Left: histogram-based density estimation of the marginal observable $x$ (polar angle) and below the learned LR validated against the histogram approach. Middle: the trained marginal LR and its gradient compared to the joint quantities used as noisy label data shown on a 1D slice and in the 2D space of observables and theory parameters. Right: Loss trajectories of LR regression with and without \madjax~gradient data.}
%  Blue points are the joint LR and
%  gradients of observed samples after detector interactions and
%  reconstruction, and orange points indicate the learned likelihood ad
%  gradients marginalized over latent variables. 
%  , where points show the samples of joint likelihoods used to learn the marginal LR. 
%\vspace{-0.6cm}
\label{fig:sbi2}
\end{center}
\vspace{-0.8cm}
\end{figure}

An important use-case for automatic gradients of MEs as provided by \madjax~is their use in simulation-based inference~\cite{Cranmer30055} (SBI) techniques that require knowledge of score $\nabla_\theta \log  p_\mathrm{ME}(z|\theta)$ variables, such as the `Mining Gold' technique~\cite{Brehmer5242} for likelihood ratio (LR) estimation. Here $r(x,\theta_0,\theta_1) = p(x|\theta_0)/p(x|\theta_1)$ is approximated by a deep neural network $\hat{r}_\phi(x,\theta_0,\theta_1)$ parametrized by weights $\phi$ whose values are optimized through regression. While the true likelihood ratio $r(x,\theta_0,\theta_1)$ is intractable, which precludes the use of standard regression objectives such as mean squared errors to the true value, it can be shown that regression to the joint likelihood ratio $r(x,z|\theta_0,\theta_1)$ yields a consistent estimator in the large-data limit. Here the joint LR acts as a ``noisy label'' which is tractable to compute. This regression on the LR can be dramatically improved by not only utilizing the joint LR, but also its gradient $t(z,x|\theta_0,\theta_1) = \nabla_\theta p(z,x|\theta)/p(z,x|\theta_1)|_{\theta_0}$, where the use of the joint values as labels yields an additional consistent training objective for the gradient $\nabla_{\theta_0 } r(x,\theta_0, \theta_1)$.
While this technique can work around the intractability of the marginalized likelihood ratio, it still requires computation of the joint quantities, in particular the gradients of the joint probabilistic model $p(x,z)$. Without an automatic solution, this approach is thus limited to cases where gradients are derivable in close-form. \madjax, can extend the method to arbitrary theory parameters. As a proof-of-concept,  Fig.~\ref{fig:sbi2} demonstrates a simplified example with the Fermi-constant as the parameter of interest, where the power of gradient-based training data is apparent from the training trajectory.

\subsection{Normalizing Flow Training using Matrix Element Gradients}

 Normalizing flows~\cite{JMLR:v22:19-1028} (NF), which enable both explicit likelihood estimation and sampling, provide a powerful method for modeling complex MEs in order to increase sampling speed and efficiency (see e.g.~\cite{PhysRevD.101.076002,10.21468/SciPostPhys.8.4.069,butter2020generative}). It can be challenging, however, to train NFs to sufficient accuracy, and training can require relatively large datasets sampled from the simulator. ME gradients provide new tools to help reduce the required scale of simulated datasets and improve modeling.

The typical maximum likelihood loss function for NF training has the form:
\begin{equation}
    \mathcal{L}_{NLL} = -\mathbb{E}_{x\sim p(x)}\Big[ \log q(x; \theta) \Big] = D_{KL}\Big[ p || q(\cdot ; \theta) \Big] + \mathrm{const.}
\end{equation}
where samples are generated by the ME simulator, $x\sim p(x)$, and the likelihood under the NF model, $q(x; \theta)$, is evaluated and minimized with respect to its parameters $\theta$.  This loss function has two challenges; (a) one must generate samples from the simulator for training, which may be computationally costly, and (b) the loss pushes the NF to be \textit{mass-distributing}, i.e. encouraging the NF to assign mass broadly so as not to assign zero density to points in which $p(x)\neq 0$.
With access to the gradient of the (un-normalized) target density, additional loss terms are available, such as the \emph{reverse KL} loss, $\mathcal{L}_{RKL}$, and a \emph{force-matching loss}, $\mathcal{L}_{FM}$, which have shown promise in lattice QCD (i.e.~\cite{PhysRevD.100.034515,sun2022finitevolume}) and molecular physics (i.e.~\cite{jaxmd2020,doi:10.1021/acscentsci.8b00913}). The losses are defined as: 
\begin{equation}
    \mathcal{L}_{RKL} = \mathbb{E}_{x\sim q(x;\theta)}\Bigg[ \log \frac{q(x; \theta)}{p(x)} \Bigg] = D_{KL}\Big[ q(x ; \theta) || p(x) \Big]
\end{equation}

\begin{equation}
    \mathcal{L}_{FM} = \mathbb{E}_{x\sim p(x)}\Bigg[ || \partial_{x} \log p(x) - \partial_{x} \log q(x; \theta) ||^{2}_{2} \Bigg]
\end{equation}
The reverse KL loss leverages much cheaper samples from the NF and exhibits \textit{mode-seeking} behavior; training under this loss pushes the NF to assign zero density where $p(x)$ is zero and hence makes it concentrate on one of the modes of $p(x)$.  The force matching pushes the NF model during training to better match the shape (as estimated with gradients) of the target ME distribution. Gradient-based training with both $\mathcal{L}_{RKL}$ and $\mathcal{L}_{FM}$ requires differentiation through the matrix element $p(x)$, and thus requires differentiable matrix elements as provided by \madjax. By mixing the forward KL with the reverse KL and/or force-matching losses, additional information about the ME is provided to the NF during training.

\begin{figure}[htbp]
    \vspace{-0.5cm}
    \centering
    \begin{subfigure}[b]{0.49\linewidth}
    \includegraphics[width=0.99\textwidth]{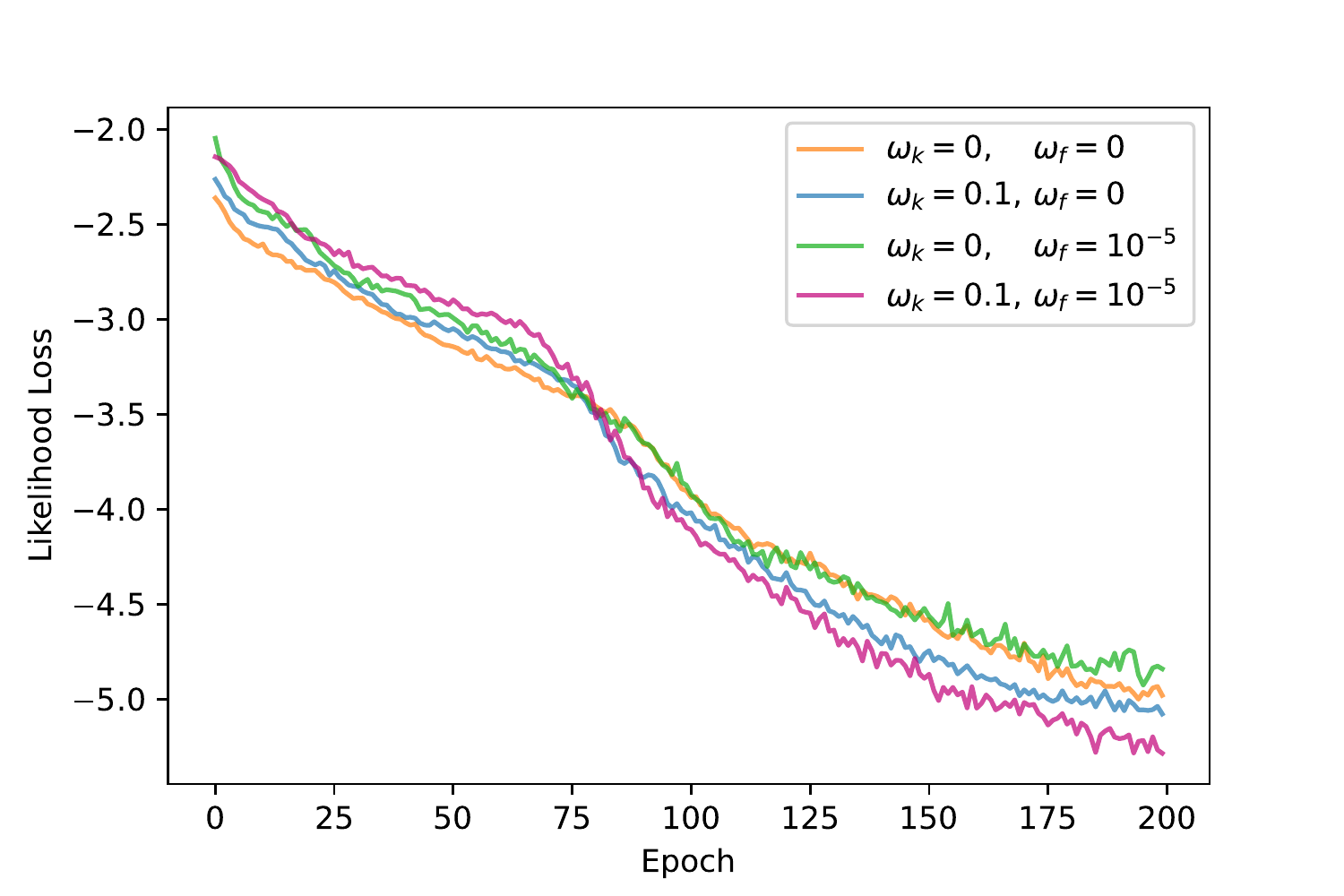}
    %\caption{ }
    \label{fig:flow_a}
    \end{subfigure}
    \hfill
     \begin{subfigure}[b]{0.49\linewidth}
    \includegraphics[width=0.99\textwidth]{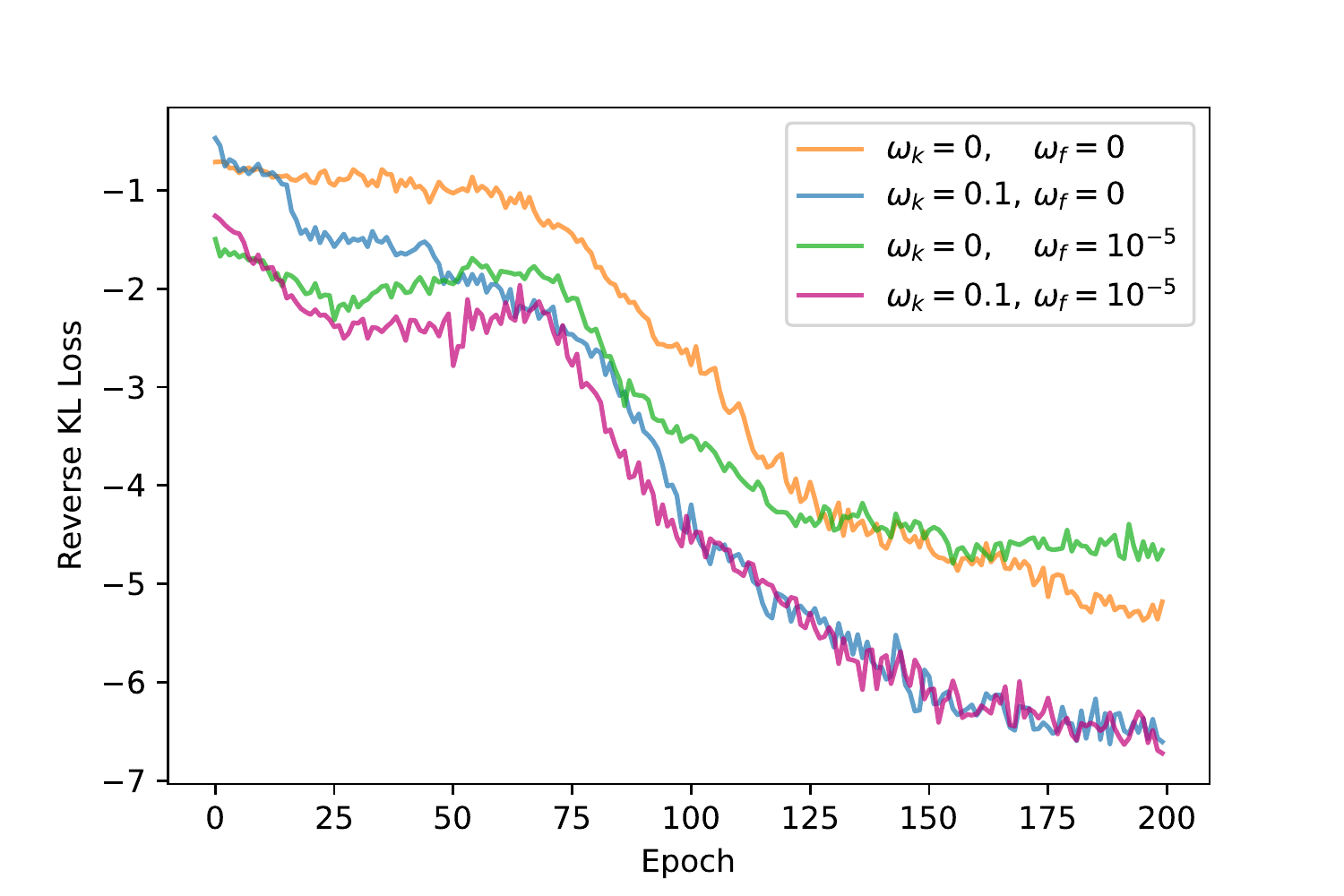}
    %\vspace{0.5cm}
    %\caption{ }
    \label{fig:flow_b}
    \end{subfigure}
     \vspace{-0.4cm}
    \caption{Comparison of training evolution of the likelihood loss (left) and  reverse KL loss (right) for models trained with and without reverse KL and force matching losses.}
    \label{fig:flow_results}
    \vspace{-0.5cm}
\end{figure}

Using \madjax, we train multiple NF models of $e^{+}e^{-}\to t\bar{t}\to q\bar{q}b q\bar{q}\bar{b}$ matrix elements. We use the loss $\mathcal{L} = \mathcal{L}_{NLL} + \omega_{k} \mathcal{L}_{RKL} +  \omega_{f} \mathcal{L}_{FM}$, with hyperparameters $\omega_{k}$ and $\omega_{f}$ controlling the impact of the reverse KL and force matching losses, respectively. The NF models comprise four layers of smooth normalizing flows~\cite{2021smooth}, which provide smooth mappings over compact intervals and thus are well adapted to mapping uniform random variables smoothly onto  ME distributions defined on phase space hypercubes. Each smooth NF layer uses four bijections and the parameters of the transformation are modeled with a MLP of two layers of 100-neurons. 8000 events are used for training using the Adam~\cite{DBLP:journals/corr/KingmaB14} optimizer with a learning rate of $5\cdot 10^{-5}$ and batch size of 100, and 2000 events are used for testing. 10 initial warm-up epochs, where all models are trained using only the $\mathcal{L}_{NLL}$, are used to improve training stability. Fig.~\ref{fig:flow_results} shows the evolution (after warm-up epochs) of the likelihood loss (left) and reverse KL loss (right), comparing models trained with and without reverse KL and force matching losses. Variations in initial model performance due to random different model initializations are overcome after epoch $\sim$75 and then models trained with $\mathcal{L}_{RKL}$ are seen to outperform the likelihood-only trained model ($\omega_k = \omega_f = 0$) in both likelihood and reverse KL. The model trained with both reverse KL and force-matching losses resulted in the best likelihood of all models. These results indicate the potential to improve ME modeling using losses enabled uniquely with  ME gradients.

\section{Conclusion}
We present \madjax, the first automatically differentiable matrix element code based on \madgraph{}. The gradient information provided by \madjax{} enables previously inaccessible approaches to HEP data analysis, such as highly data-efficient methods for training ML models for high-energy physics or simulation-based inference. \madjax{} is a prototype that serves as proof-of-principle of the utility of such tools for future R\&D. The implementation of \madjax{} using \jax{} facilitates easy use with ML pipelines, but \jax~is not always well-adapted to the structure of matrix element code. Future directions will explore the use of alternative AD tools that may be better suited to matrix element code structures.

\section*{Acknowledgements}
We thank Valentin Hirschi for providing the initial python-\madgraph{} interface code and thank Matthew Feickert for the software guidance, and thank both for the helpful discussions. We thank Andreas Kr\"amer and Jonas Koehler for supplying their code for Smooth Normalizing Flows. 

\noindent MK is supported by the US Department of Energy (DOE) under grant DE-AC02-76SF00515.

\noindent LH is supported by the Excellence Cluster ORIGINS, which is funded by the Deutsche Forschungsgemeinschaft (DFG, German Research Foundation) under Germany’s Excellence Strategy - EXC-2094-390783311.

\section*{References}
\bibliographystyle{iopart-num}
\bibliography{bibliography}

\end{document}